# Weak Ferromagnetism in LaCo$_{1-x}$Rh$_x$O$_3$: Anomalous Magnetism Emerging between Two Nonmagnetic End Phases

Shinichiro Asai, Noriyasu Furuta, Yukio Yasui and Ichiro Terasaki

Department of Physics, Nagoya University, Furo-cho, Chikusa-ku, Nagoya 464-8602

Abstract

Magnetization has been measured for polycrystalline samples of LaCo$_{1-x}$Rh$_x$O$_3$ ($0 \leq x \leq 0.9$) in order to investigate magnetism induced in the solid solution of two nonmagnetic end phases of LaCoO$_3$ and LaRhO$_3$. It is found that a ferromagnetic transition is observed below 15 K in the range of $x$ from 0.1 to 0.4. The effective Bohr magnetic moment evaluated from the temperature dependence of magnetic susceptibility at around room temperature is independent of $x$ for $0 < x < 0.5$ (approximately 3 $\mu_B$ per formula unit), and rapidly decreases above $x = 0.5$. On the basis of detailed magnetization measurements, the spin state and magnetic ordering of LaCo$_{1-x}$Rh$_x$O$_3$ are discussed.

1. Introduction

The perovskite cobalt oxide $LaCoO_3$ is a fascinating material, because the spin state of $Co^{3+}$ ions changes with temperature and pressure.[1-6] In this oxide, the $Co^{3+}$ ($3d^6$) ion is surrounded with six octahedrally-coordinated $O^{2-}$ ions, and the 3d orbitals split into the $e_g$ orbitals in the higher energy level and the $t_{2g}$ orbitals in the lower energy level (crystal field splitting). The $Co^{3+}$ ion takes the high-spin state (HS: $e_g^2 t_{2g}^4$, $S = 2$), when the Hund coupling is larger than the crystal field splitting, while it does the low-spin state (LS: $t_{2g}^6$, $S = 0$) for the opposite condition. In addition, the intermediate-spin state (IS: $e_g^1 t_{2g}^5$, $S = 1$) has been proposed by introducing a finite transfer between the $e_g$ orbitals and the O 2p orbitals.[7] The ground state of $Co^{3+}$ in $LaCoO_3$ is known as the low-spin state below 100 K,[3] but excited states are controversial at present, though many researchers have experimentally and theoretically studied this issue for many years.[7-10]

$LaCo_{1-x}M_xO_3$ shows various anomalous behavior because the substituted M ion affects the spin state of $Co^{3+}$ nearby.[11] In particular, when M = Ga ($3d^{10}$), Al ($3d^0$), and Rh ($4d^6$), $LaCo_{1-x}M_xO_3$ at low temperature can be regarded as a solid solution of nonmagnetic phases. Magnetic properties of $LaCo_{1-x}M_xO_3$ (M = Ga, Al, Rh) were reported by Kyomen et al.[12] $LaCo_{1-x}Ga_xO_3$ and $LaCo_{1-x}Al_xO_3$ are nonmagnetic at low temperature, indicating that $Ga^{3+}$ and $Al^{3+}$ substitution leaves the low-spin state stable. On the contrary, in the case of $LaCo_{1-x}Rh_xO_3$, a Curie-Weiss-like susceptibility is developed with $Rh^{3+}$ substitution, and the spin-state crossover to the low-spin state seems to disappear for $x > 0.04$. This behavior is highly nontrivial, and an extra effect and/or a cooperative phenomenon may occur in $LaCo_{1-x}Rh_xO_3$. Rh substitution for Co causes anomalous physical properties in other cobalt oxides; An ordered structure of the high-spin state $Co^{2+}$ and the low-spin state $Rh^{4+}$ is reported in $Ca_3CoRhO_6$ with one-dimensional chains.[13] The Rh substitution for Co in the misfit cobalt oxides $[Pb_{0.7}Co_{0.4}Sr_{1.9}O_3][CoO_2]_{1.8}$ and $Bi_{1.7}Ba_2Co_2O_\alpha$ does not affect the thermopower, although it increases the electric resistivity.[14,15] For $La_{0.8}Sr_{0.2}CoO_3$, Rh substitution for Co anomalously increases the Seebeck coefficient.[16] These anomalous behaviors should be related to the magnetism of $LaCo_{1-x}Rh_xO_3$.

Recently, Li et al. reported the thermoelectric and structural properties of $LaCo_{1-x}Rh_xO_3$, and proposed that a Rh ion substituted for Co exists as tetravalent, and makes one Co ion divalent.[17] Because of this complexity, they did not analyze the magnetic susceptibility of $LaCo_{1-x}Rh_xO_3$. In this paper, we have measured the x-ray diffraction and magnetization of $LaCo_{1-x}Rh_xO_3$ in the range of $0 \leq x \leq 0.9$, and have found a ferromagnetic ordering in the range of $0.1 \leq x \leq 0.4$. We emphasize that the ferromagnetic ordering induced by $Rh^{3+}$ substitution is newly discovered, and is distinguished from the metallic ferromagnetic state driven by the double exchange mechanism, as is seen in $La_{1-x}Sr_xCoO_3$.[18] This result indicates that the magnetism of $LaCo_{1-x}Rh_xO_3$ should be understood as a cooperative phenomenon. We evaluate the effective Bohr magnetic moment by fitting the susceptibility with the Curie-Weiss law near room temperature, and have found that the effective Bohr magnetic moment is independent of $x$ for $0 \leq x \leq 0.5$, and rapidly decreases above $x = 0.5$. Combining these results with the measured lattice volume, we discuss the spin state of $LaCo_{1-x}Rh_xO_3$ at room temperature.

2. Experiments

Polycrystalline samples of $LaCo_{1-x}Rh_xO_3$ ($0 \leq x \leq 0.9$) were prepared by a conventional solid-state reaction. Mixture of $La_2O_3$ (3N), $Co_3O_4$ (3N) and $Rh_2O_3$ (3N) with stoichiometric molar ratios was ground, and calcined for 24 h at 1000°C in air. The calcined powder was ground, pressed into a pellet, and sintered for 48 h at 1200°C in air.

X-ray diffraction was measured with a Rigaku Geigerflex ($CuK_\alpha$ radiation), and no impurity phases were detected for the prepared samples. The lattice parameter obtained from x-ray patterns is consistent with the reported value in the literature.[17] Magnetization $M$ in field cooling (FC) and zero field cooling (ZFC) processes was measured using SQUID magnetometer (Quantum Design MPMS) from 5 to 300 K in an applied field $H$ of 1 T for $0 \leq x \leq 0.9$. For $0.1 \leq x \leq 0.5$, low-field magnetization was also measured in 0.05 T. Magnetization-field ($M$ - $H$) curves for $x = 0.2$ were measured in sweeping $H$ from 0 to 7 T, and from 7 down to 0 T.

3. Results and Discussion

Figure 1(a) shows the $x$ dependence of the lattice volume of $LaCo_{1-x}Rh_xO_3$ per formula unit (f. u.) at room temperature. Although the crystal structure changes from rhombohedral to orthorhombic at $x = 0.2$, the lattice volume continuously increases with $x$, which is consistent with the results reported by Li et al.[17] We should note that the lattice volume is larger than that expected from Vegard's law, as is shown by the broken line in Fig. 1(a). A model for the larger volume represented by the solid line is discussed later. The $x$ dependence of the lattice constant is shown in Fig. 1(b), where the lattice parameter $a$ increases continuously with increasing $x$. The angle of rhombohedral α increases from 60.8 to 60.9 degree as the Rh content increases from $x = 0$ to $x = 0.1$ (not shown).

The temperature dependence of magnetization per formula unit taken in 1 T for

LaCo$_{1-x}$Rh$_x$O$_3$ is shown in Figs. 2(a) and 2(b). For $x = 0$, the magnetization drops below 100 K, which has been understood as the spin-state crossover. In contrast, for $x = 0.1$, the magnetization increases with decreasing temperature which indicates that the spin-state crossover disappears. The magnetization takes a maximum at $x = 0.2$ as a function of $x$ below 150 K. Figure 3(a) shows the magnetization of $x = 0.2$ taken in 0.05 T, where a hysteresis between FC and ZFC is clearly observed below 10 K. A similar hysteresis was also observed in $x = 0.1$, 0.3, and 0.4 (not shown). Figure 4 shows the $M$ - $H$ curves for $x = 0.2$. One can clearly see a magnetic hysteresis at 2 K with a spontaneous magnetization, which indicates a ferromagnetic order at this temperature. The ferromagnetic transition temperature $T_C$ is defined by the temperature at which $|\,\mathrm{d}M\,/\,\mathrm{d}T\,|$ takes a maximum value of the magnetization in 0.05 T (inflection point of the $M$ - $T$ curve), as shown in Fig. 3(b). Figure 3(c) shows the $x$ dependence of $T_C$, with increasing Rh content, the ferromagnetic transition appears at $x \sim 0.1$, and the system has a maximum value of $T_C$ at $x = 0.2$. $T_C$ decreases for $x > 0.2$ and finally disappears at $x \sim 0.5$. It should be emphasized that the weak ferromagnetic ordering is induced without carrier dopng. As is well-known, carrier doping induces the ferromagnetic and metallic state through the double exchange interaction, as is seen in La$_{1-x}$Sr$_x$CoO$_3$.[18] However, for LaCo$_{1-x}$Rh$_x$O$_3$, the system remains as an insulating phase, when the Rh ions substitute for Co ones. Thus, the ferromagnetism of LaCo$_{1-x}$Rh$_x$O$_3$ is distinguished from that of La$_{1-x}$Sr$_x$CoO$_3$.

Note that the magnetization does not saturate and increases linearly above 6 T as shown in Fig. 4, which suggests that this magnetic order is not a simple ferromagnetic order. We estimate the saturation magnetization $M_S$ and the high-field susceptibility $\Delta M\,/\,\Delta H$ by fitting the $M$ - $H$ curve linearly above 6 T. As shown by the dotted line in Fig. 4, $M_S$ and $\Delta M\,/\,\Delta H$ are given by the intersection and slope of the fitting line, respectively. Figure 5(a) shows the temperature dependence of the saturation magnetization, which gradually increases with decreasing temperature below around 40 K. Because the spontaneous magnetization is obtained from the data above 6 T, it includes a ferromagnetic component enhanced by the applied magnetic field, and is somewhat overestimated. Figure 5(b) shows $\Delta M\,/\,\Delta H$ from 2 to 50 K. With decreasing temperature, $\Delta M\,/\,\Delta H$ drops below 40 K, indicating that an antiferromagnetic component increases and/or a paramagnetic component decreases with growing ferromagnetic component. Considering that the spontaneous magnetization at low temperature is much smaller than the effective Bohr magnetic moment at room temperature (see below), we think that the paramagnetic and/or antiferromagnetic component overlaps the ferromagnetic component. These features are what one can see in a ferrimagnet or a canted antiferromagnet. In this context, this ordering should be understood as weak ferromagnetism.

We analyze the $H/M$ - $T$ curve from 200 to 300 K by the Curie-Weiss law

$$\frac{H}{M} = \frac{3k_B}{\mu_{eff}^2 \mu_B}(T+\theta), \qquad (1)$$

and evaluate the effective Bohr magnetic moment $\mu_{eff}$ and the Curie-Weiss temperature $\theta$. Figure 6 shows the $x$ dependence of $\mu_{eff}$, where the values of $\mu_{eff}$ are close to 3 $\mu_B$ below $x = 0.5$, and rapidly decrease above $x = 0.5$. The values of $\theta$ are positive and between 80 and 300 K (not shown), indicating that $LaCo_{1-x}Rh_xO_3$ has an antiferromagnetic interaction at room temperature.

Here, we discuss an origin of the magnetism of $LaCo_{1-x}Rh_xO_3$. As for the excited state of $LaCoO_3$ at room temperature, the following two models are proposed.[7-10] One is that all the $Co^{3+}$ ions are in the intermediate-spin state (model 1), while the other model is that about a half of the $Co^{3+}$ ions are in the high-spin state, and the rest of them remain in the low spin state (model 2). If the model 1 were valid, we would naturally expect the spin state of $LaCo_{1-x}Rh_xO_3$ to be (1 - $x$) IS $Co^{3+}$ + $x$ $Rh^{3+}$ and therefore

$$\mu_{eff} = 2\sqrt{(1-x)(2-x)}, \qquad (2)$$

as shown by the broken line in Fig. 6. Obviously this is incompatible with the experimental results. On the contrary, as for the model 2, the effective Bohr magnetic moment of $LaCo_{1-x}Rh_xO_3$ can be independent of $x$ below $x = 0.5$, if we assume that the nonmagnetic $Rh^{3+}$ ion selectively substitutes for the low-spin state $Co^{3+}$ ion. This assumption is reasonable; the ionic radius of $Rh^{3+}$ (0.665 Å)[19] is larger than that of the low-spin state $Co^{3+}$ (0.545 Å)[19], which is consistent with the $x$ dependence of the lattice volume shown in Fig. 1. We further assume that the $Rh^{3+}$ ion starts to replace the high-spin state $Co^{3+}$ ion (0.61 Å)[19] for $x > 0.5$. The expected volume and $\mu_{eff}$ of this model are shown by the solid curves in Figs. 1 and 6, which reasonably explain the observed $x$ dependence of the experimental data.

Next we examine a possibility that the substituted Rh ion exists as tetravalent ($Rh^{4+}$), and concomitantly creates one $Co^{2+}$ ion, as was suggested by Li et al.[17] Assuming the high-spin state $Co^{2+}$ (S = 3/2) and the low-spin state $Rh^{4+}$ (S = 1/2), we estimate $\mu_{eff}$ to be close to $\mu_{eff}$ of model 2. In this respect, we cannot exclude this possibility of the magnetic susceptibility near room temperature. When the orbital moment survives in $Co^{2+}$ as was reported in CoO,[20] the total moment can be larger than S = 3/2. In this case $\mu_{eff}$ increases with $x$ for small $x$, which is obviously incompatible with the experimental data in Fig. 6. The lattice volume shown in Fig.1 is favorable to model 2 rather than the

$Co^{2+}$-$Rh^{4+}$ pair creation. When the high-spin state $Co^{2+}$ ions are created by Rh substitution, the lattice volume is expected to have a larger value than the solid line in Fig. 1, because ionic radius of the high-spin state $Co^{2+}$ (0.745 Å) [19] is considerably larger than those of the low-spin state $Rh^{3+}$ (0.665 Å) [19].

Finally let us comment on the ground state of $LaCo_{1-x}Rh_xO_3$. As shown in Fig. 2, the spin state crossover disappears in $LaCo_{1-x}Rh_xO_3$, indicating that $Rh^{3+}$ stabilizes the neighboring high-spin state $Co^{3+}$ down to low temperature. Then, we naturally expect that the high-spin state $Co^{3+}$ is responsible for the ferromagnetic ordering of $LaCo_{1-x}Rh_xO_3$. If the $Rh^{3+}$ substitution effect were simply to expand the lattice, the ferromagnetism observed here would be compared with the ferromagnetism in a strained film of $LaCoO_3$.[21, 22] However, we notice that the $Ga^{3+}$ substitution for $Co^{3+}$ in $LaCoO_3$ also expands the lattice volume,[23] but $LaCo_{1-x}Ga_xO_3$ is nonmagnetic at low temperature.[12] We should emphasize that the lattice volumes of $LaCo_{0.6}Ga_{0.4}O_3$ and $LaCo_{0.9}Rh_{0.1}O_3$ are nearly equal, but the low-temperature magnetism is entirely different. In this context, we speculate that $Rh^{3+}$ leaves high-spin state $Co^{3+}$ stable by some extra effect (for example, an interaction between the $Co^{3+}$ 3d and $Rh^{3+}$ 4d orbitals). The substituted $Rh^{3+}$ ion plays two roles in this ordering; one is to keep the neighboring $Co^{3+}$ ions magnetic, and the other is to disorder the ferromagnetism caused by the $Co^{3+}$ ions. The latter effect is supported by the result that the Curie-Weiss temperature evaluated at room temperature is much higher than the ferromagnetic transition temperature. We suggest that this disordering effect suppresses the saturation magnetization, and complicates the magnetic structure of $LaCo_{1-x}Rh_xO_3$. Detailed magnetic structure should be clarified, by means of neutron measurements using single-crystalline sample of $x = 0.2$, which are in progress.

4. Summary

We have prepared a set of polycrystalline samples of $LaCo_{1-x}Rh_xO_3$ ($0 \leq x \leq 0.9$) and have measured the magnetization. The effective Bohr magnetic moments are 3 $\mu_B$ per formula unit, nearly independent of $x$ below $x = 0.5$, and rapidly decrease with increasing $x$ above $x = 0.5$. This result suggests that the nonmagnetic $Rh^{3+}$ is substituted for the low-spin state $Co^{3+}$ selectively, and implies that the spin state of $Co^{3+}$ in $LaCo_{1-x}Rh_xO_3$ at room temperature can be regarded as a mixture of the high- and low-spin states. This is quantitatively consistent with the $x$ dependence of the lattice volume. We have also found a magnetic hysteresis at low temperature in $LaCo_{1-x}Rh_xO_3$ in the range of $0.1 \leq x \leq 0.4$, which indicates weak ferromagnetism. The spontaneous magnetization at low temperature is smaller than the effective Bohr magnetic moment evaluated at room temperature, and the magnetization continues to increase linearly above 6 T. These results suggest that the observed weak ferromagnetism comes from ferrimagnetism or canted antiferromagnetism. This weak ferromagnetism is unique in the sense that a magnetically ordered state arises from a mixture of nonmagnetic materials.

Acknowledgment

We would like to thank S. Shibasaki for preliminary study of this system, and R. Okazaki for fruitful discussion. This work is partially supported by Grant-in-Aid for Scientific Research, MEXT (No. 22014005), and by Strategic Japanese-Finland Cooperative Program, JST.

Figure captions

Fig.1. (a) The lattice volume per formula unit of $LaCo_{1-x}Rh_xO_3$ plotted as a function of the Rh content $x$ ($0 \leq x \leq 1$). Broken and solid lines represent the lattice volumes expected from Vegard's law and from our proposed models (see text) respectively. (b) The lattice parameter of $LaCo_{1-x}Rh_xO_3$ plotted as a function of the Rh content $x$ ($0 \leq x \leq 1$).

Fig.2. Temperature dependence of magnetization of $LaCo_{1-x}Rh_xO_3$ taken in 1 T for (a) $0 \leq x \leq 0.2$, and (b) $0.3 \leq x \leq 0.9$ respectively.

Fig.3. (a) Temperature dependence of magnetization for $x = 0.2$ measured in 0.05 T. (b) Temperature dependence of $|dM/dT|$ in field cooling for $0.2 \leq x \leq 0.5$ measured in 0.05 T. (c) $x$ dependence of the transition temperature for $LaCo_{1-x}Rh_xO_3$ ($0 \leq x \leq 0.5$) determined by the temperature at which $|dM/dT|$ takes a maximum.

Fig.4. (Color online) Magnetization-field ($M - H$) curve for $x = 0.2$. The dotted line represents the linear fit of the data above 6 T (see Fig. 5).

Fig.5. The results of the linear fit of the high-field magnetization for $x = 0.2$ (see text). (a) the saturation magnetic moment $M_S$, and (b) the magnetic susceptibility $\Delta M / \Delta H$.

Fig.6. The effective Bohr magnetic moment $\mu_{eff}$ plotted as a function of Rh content $x$. The Broken and solid curves show calculation from two different models (see text).

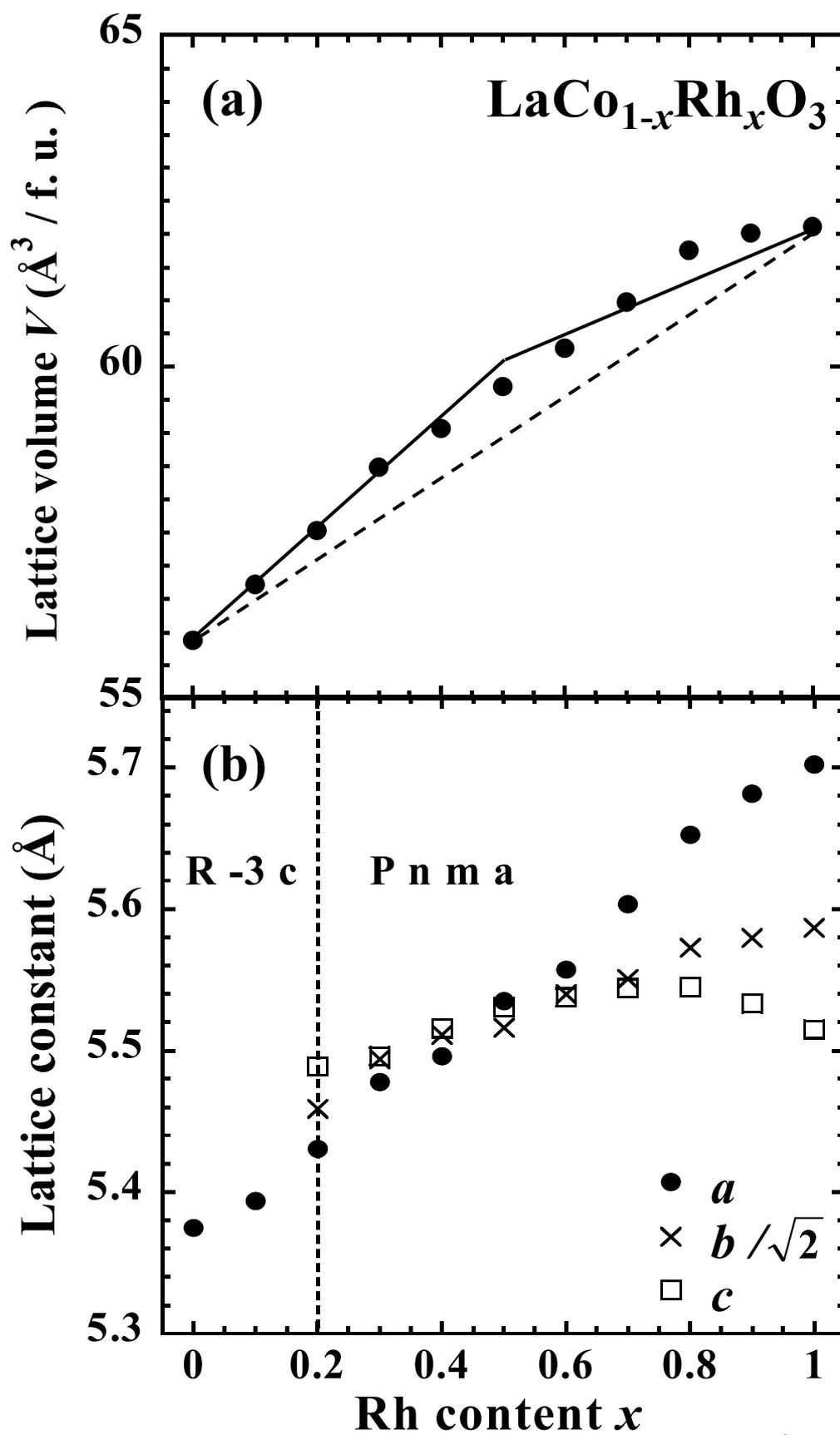

Fig. 1
S. Asai et al.

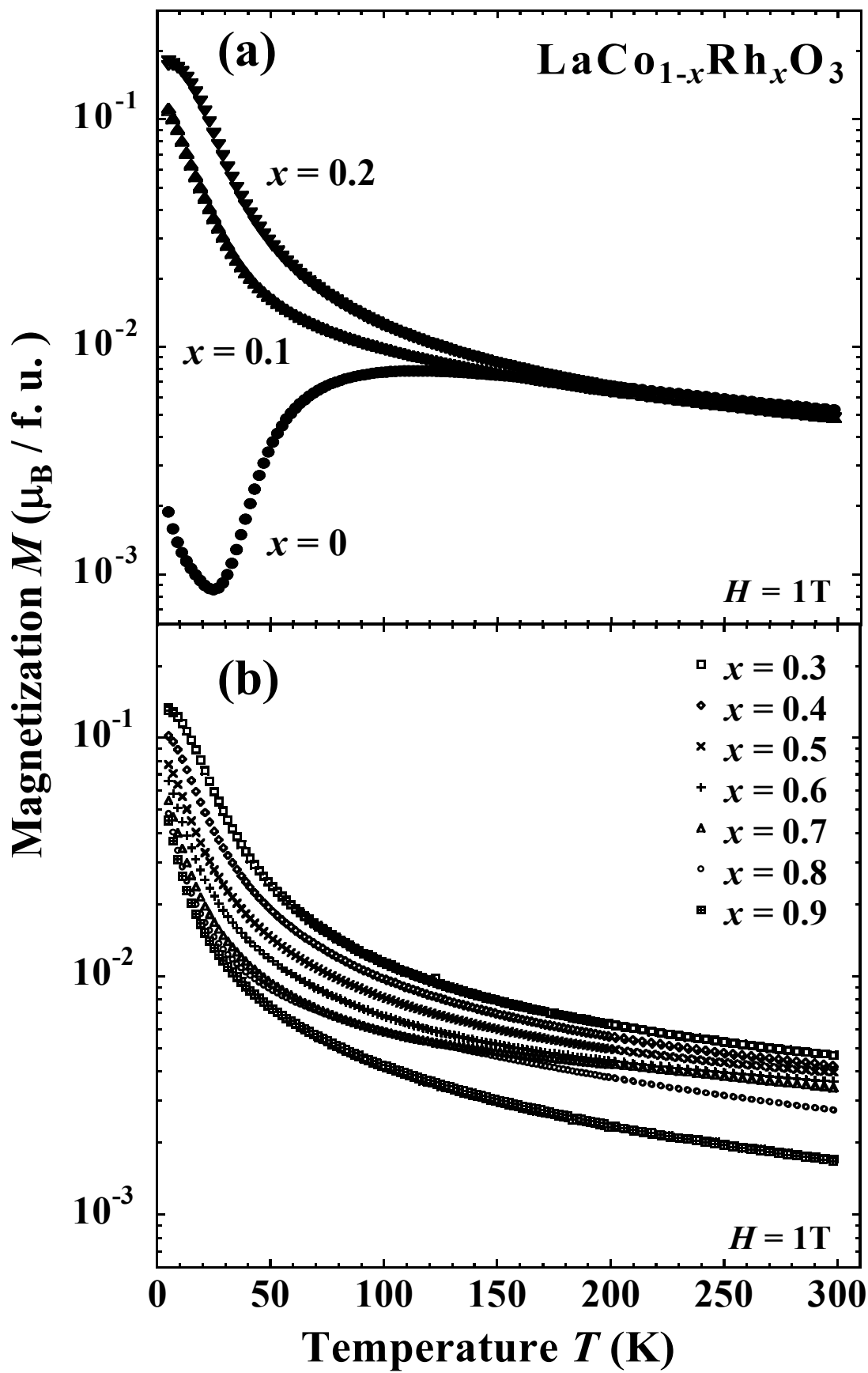

Fig. 2
S. Asai et al.

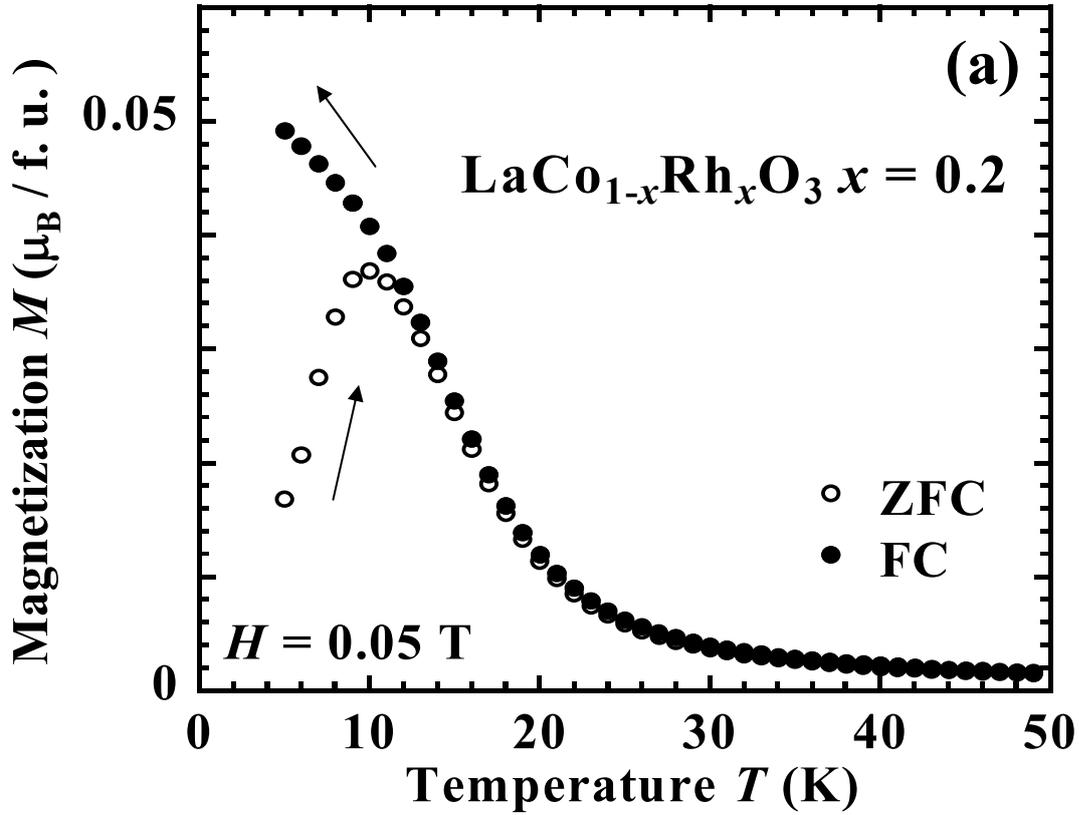
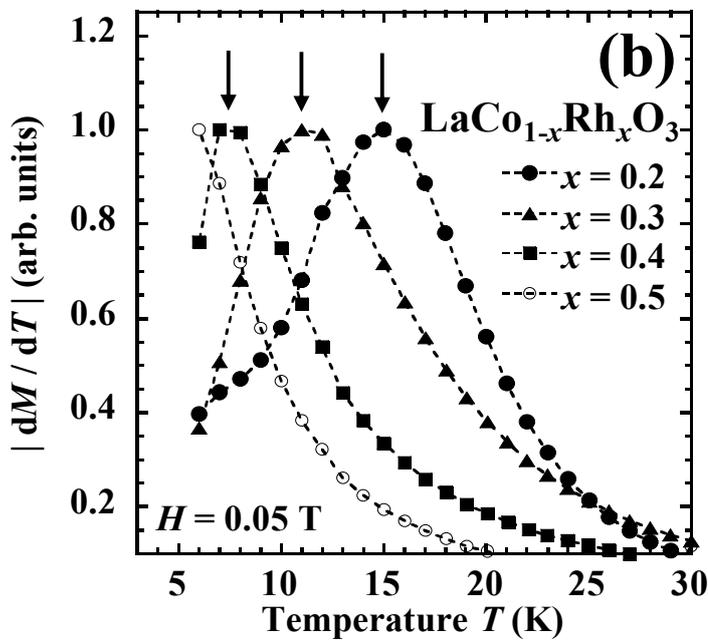
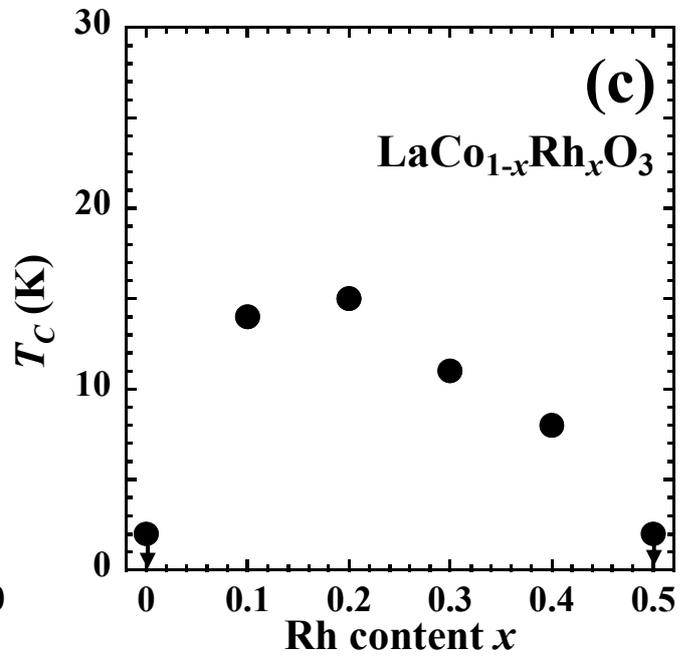

Fig. 3
S. Asai *et al.*

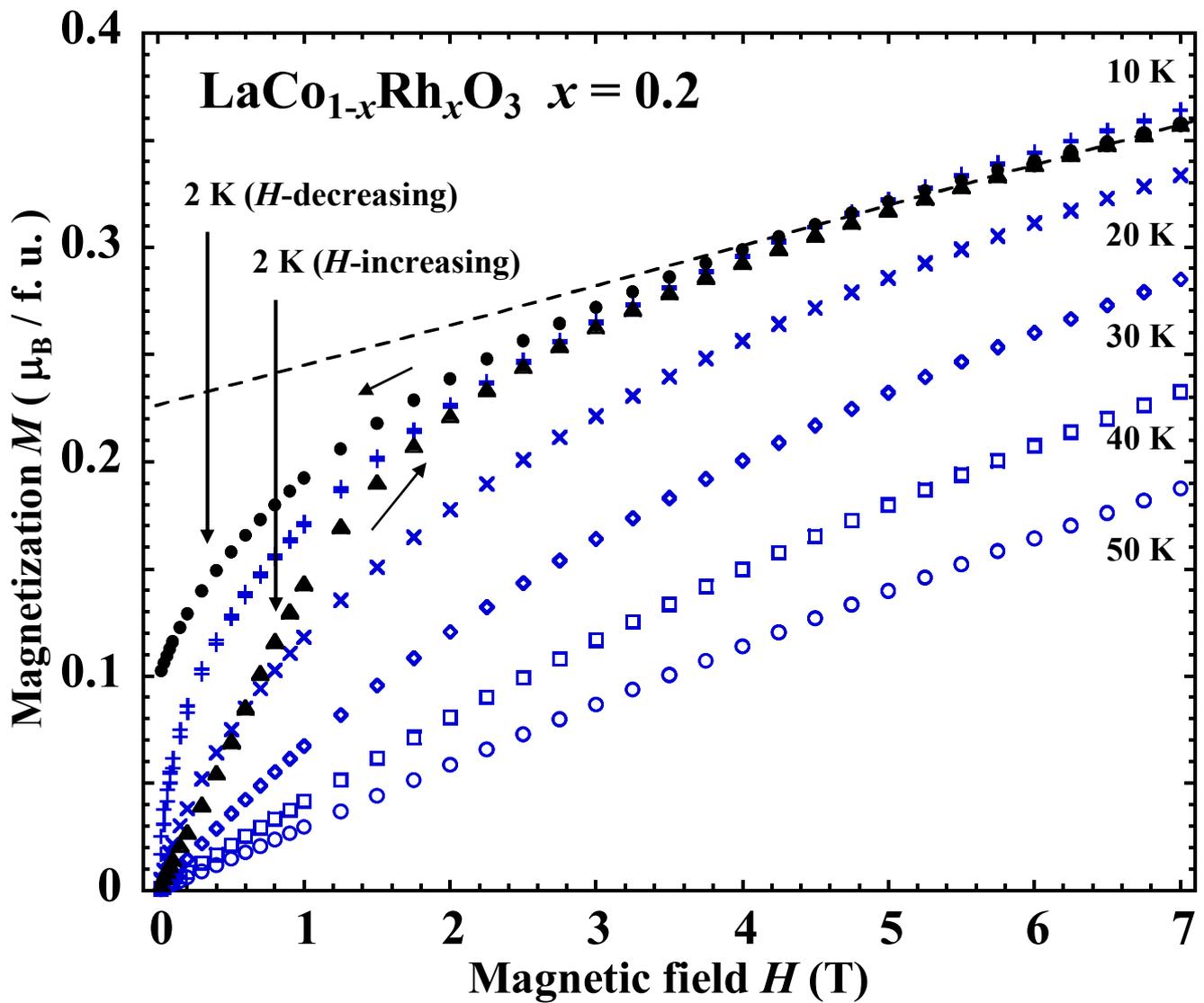

Fig. 4
S. Asai *et al.*

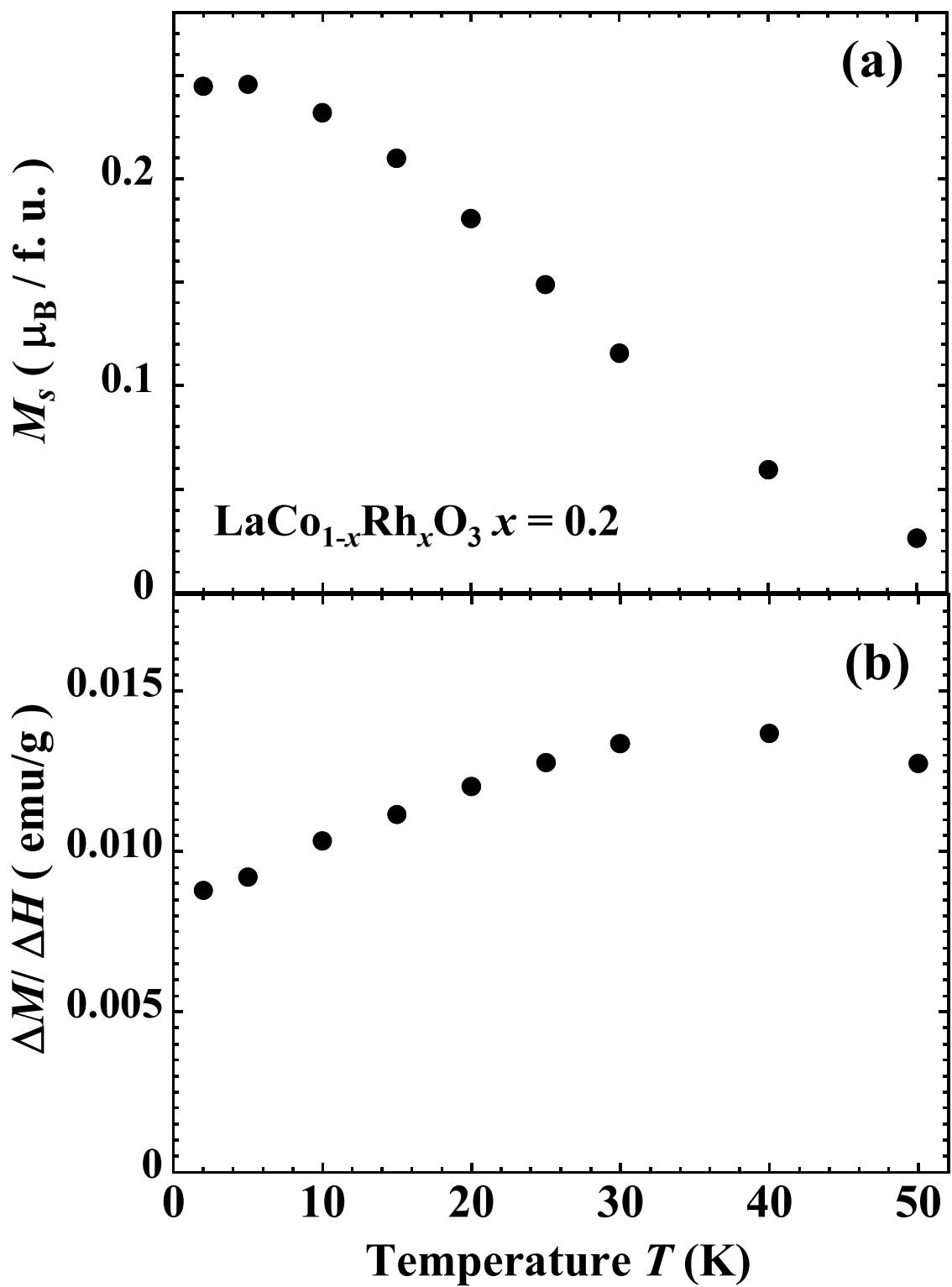

Fig. 5
S. Asai *et al.*

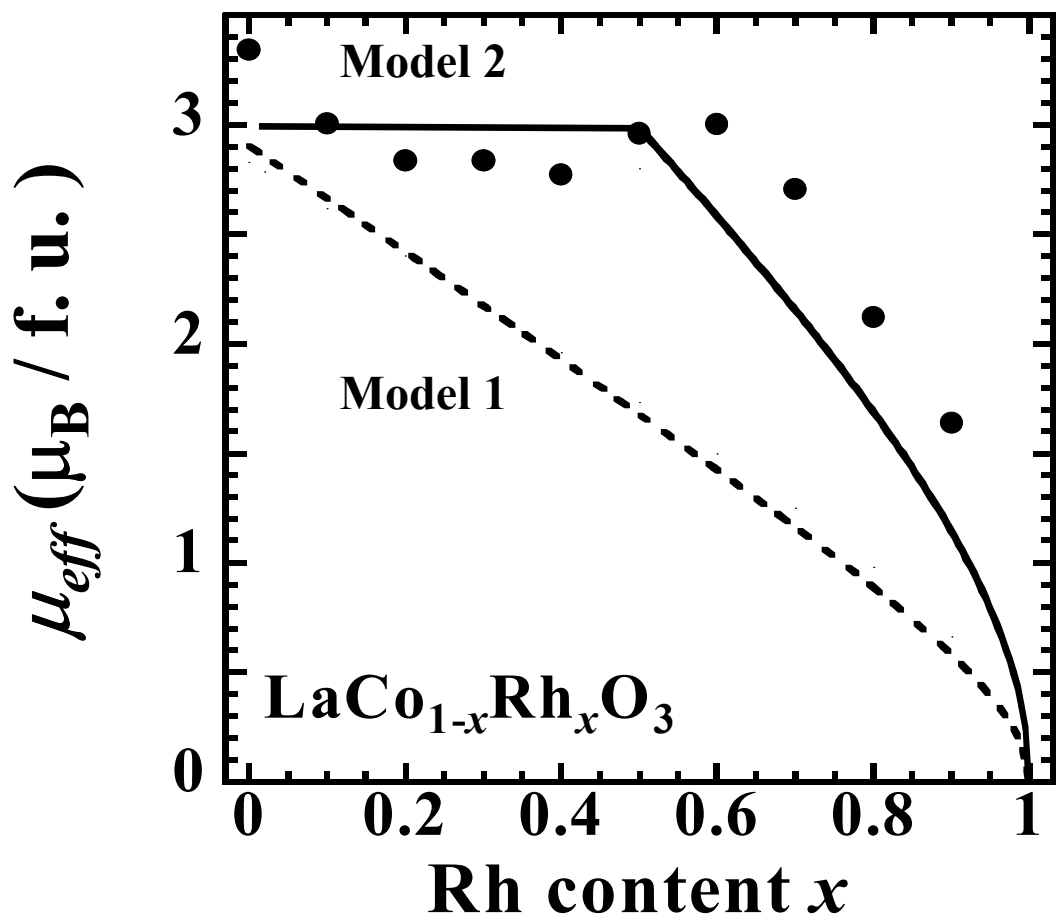

Fig. 6
S. Asai *et al.*